\documentclass[fleqn,12pt,twoside]{article}
\usepackage{espcrc1}

\usepackage{graphicx}

\usepackage[figuresright]{rotating}

\newcommand{\AmS}{{\protect\the\textfont2
  A\kern-.1667em\lower.5ex\hbox{M}\kern-.125emS}}

\title{Directed and Elliptic Flow in Pb+Pb collisions at 40 and 158 AGeV}

\author{Alexander Wetzler$^{5}$, N.~Borghini$^{2}$, P.M.~Dinh$^{6}$, J.-Y.~Ollitrault$^{6,7}$, A.M.~Poskanzer$^{1}$, S.A.~Voloshin$^{3}$ and the NA49-collaboration$^*$
\vspace{0.4cm}
\begin{small}
\newline
$^{1}$Lawrence Berkeley National Laboratory, Berkeley, CA, USA.\\
$^{2}$Universit{\'e} Libre de Bruxelles, Brussels, Belgium.\\
$^{3}$Wayne State University, Detroit, MI, USA.\\
$^{5}$Fachbereich Physik der Universit\"{a}t, Frankfurt, Germany.\\
$^{6}$CEA-Saclay, Gif-sur-Yvette, France.\\
$^{7}$L.P.N.H.E., Universit{\'e} Pierre et Marie Curie, Paris, France.\\
$^*$For complete NA49 author list see C.~Blume or M.~v.~Leeuwen, these proceedings.\\
\end{small}
}
\begin{document}
\maketitle

\begin{abstract}
Directed and elliptic flow are reported for charged pions and protons as a function of transverse momentum, rapidity, and centrality in 40 and 158 AGeV Pb + Pb collisions. The standard method of correlating particles with an event plane is used.
The directed flow of protons is small and shows little variation near to midrapidity, but rises fast towards projectile rapidity in the 40~AGeV data. For most peripheral collisions the flat region becomes negative resulting in $v_1$ changing sign three times.
Elliptic flow doesn't seem to change very much from 40~AGeV to 158~AGeV. The difference is smaller than anticipated from the overall energy dependence from AGS to RHIC.
\end{abstract}

\section{Introduction}
Directed and elliptic flow is sensitive to the equation of state of nuclear matter. As the energy density in the interaction zone varies with collision energy and impact parameter, the energy and centrality dependence can provide information about the equation of state and a possible phase transition from hadronic to partonic matter.

\section{Data and Analysis}
Directed and elliptic flow measurements at the full beam energy of 158~AGeV were published by NA49 \cite{Pos99} based on the statistics of 120k events. All together NA49 recorded in 1996 360k events using a minimum bias centrality trigger. To increase statistics in the central bins we added 120k events taken with a trigger selecting only the 12.5\% most central events. In 1999 additional data were taken for the reduced beam energy of 40~AGeV. At this energy we have a total of 730k minimum bias events.
In the figures we are using three different centrality selections. The 12.5\% most central collisions are labeled as central, the centrality 12.5\%-33.5\% as mid-central and 33.5\%-100\% as peripheral.

An event plane method was applied to reconstruct the flow \cite{Pos98}. For the $v_2$ event plane determination a $p_t$-weight  was used to reduce the influence of non-flow two-particle correlations for particles with low transverse momentum. For the $v_1$ event plane determination the rapidity in the center of mass system was used as a weight. We also added a correction for correlations caused by transverse momentum conservation \cite{Bor03}.
We also applied a multi-particle correlation cumulant method to the data. Some results are shown in \cite{Bor02}.

\section{Results}

\begin{figure}[h]\begin{center}\includegraphics[scale=.75]{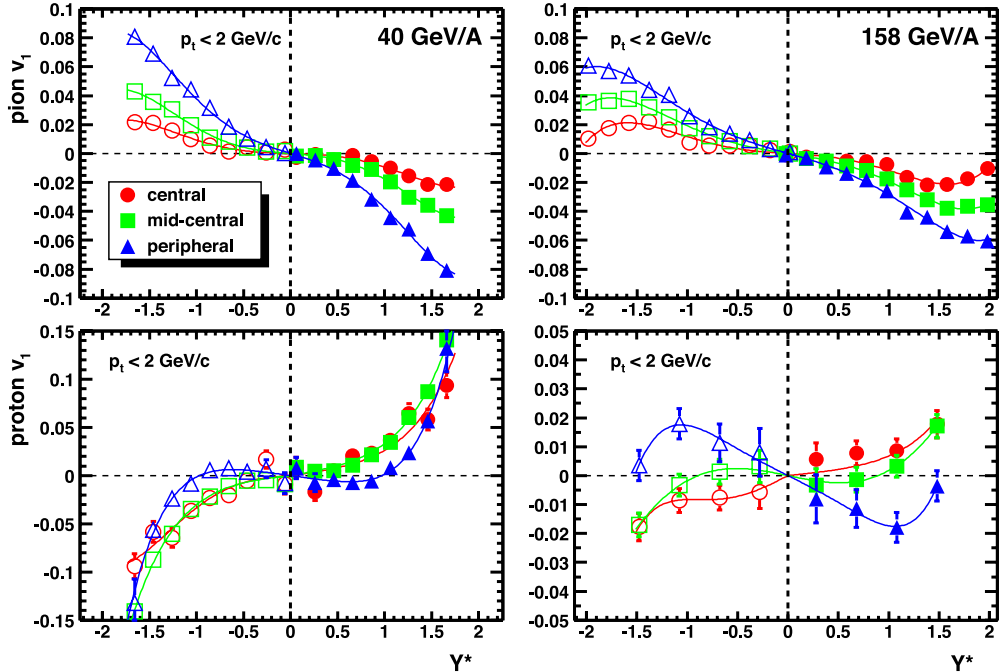}\end{center}\vspace{-1.0cm}\caption{Pion (top) and proton (bottom) directed flow at 40~AGeV (left) and 158~AGeV (right). Open points are reflected at mid-rapidity. Please note the different scale on the the two proton plots. The lines are there to guide the eye.}\vspace{-0.5cm}\label{v1graph}\end{figure}

The dependence of directed flow, $v_{1}$, on rapidity is shown in Fig.~\ref{v1graph} for 40 and 158~AGeV data and for pions and protons. At 40~AGeV the sign of the proton directed flow is taken to be positive approaching projectile rapidity. This determines the pion flow to be negative in this region. At 158~AGeV the proton data do not approach close enough to projectile rapidity to get a clear signal, so we take the pion flow to be negative in this region as at the lower energy.
As required by symmetry directed flow of pions at 40~AGeV vanishes at midrapidity. Its magnitude increases towards projectile and target and it increases from central to peripheral collisions.
Directed flow of pions at 158~AGeV shows qualitatively the same behaviour as for 40~AGeV.
At 40~AGeV for protons there is a quite flat plateau around midrapidity. About one unit of rapidity away from midrapidity it increases rapidly. At 158~AGeV it seems to be positive for central collisions and changes to negative sign for peripheral collisions. Comparing it to 40~AGeV one has to take into account the different acceptance for 40 and 158~AGeV: At 40~AGeV the rapidity coverage relative to the rapidity gap is larger than at 158~AGeV. So the shape might be similar for both energies, but the strong rise at the lower energy cannot be observed at the full energy. Assuming the same shape for both energies, results in a shape for peripheral collisions, which crosses the zero line three times \cite{Sne00}. This {\it wiggle} shape is also visible at 40~AGeV, but not as pronounced. However non flow correlations due to resonance decays, which can be of same order of magnitude, might be the reason for this changing sign \cite{Oll00,Oll01}.

\begin{figure}[h]\begin{center}\includegraphics[scale=.75]{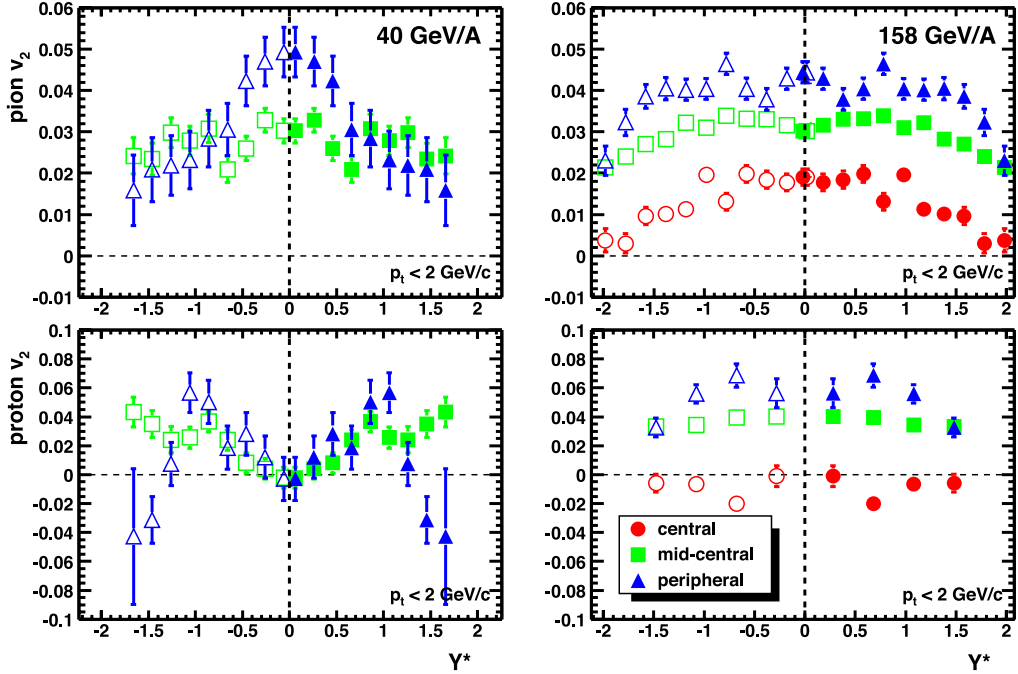}\end{center}\vspace{-1.0cm}\caption{Pion (top) and proton (bottom) elliptic flow at 40~AGeV (left) and 158~AGeV (right). Open points are reflected at mid-rapidity.}\label{v2graph}\vspace{-0.5cm}\end{figure}

In Figure~\ref{v2graph} the rapidity dependence of the elliptic flow, $v_{2}$, of pions and protons is presented for three different centrality selections.
For pions it is quite constant in rapidity for mid-central collisions and peaks near midrapidity for peripheral collisions. For most central collisions $v_2$ was not measured, because the flow is presumably very small, which results in very large statistical errors because of the poor event plane resolution. Elliptic flow of protons seems to be minimal near midrapidity and increase near projectile and target rapidity.
In contrast at 158~AGeV the rapidity dependence of pion elliptic flow doesn't change shape very much with centrality but its average value increases for more peripheral collisions. The same is observed for proton elliptic flow.

\begin{figure}\begin{center}\includegraphics[scale=.75]{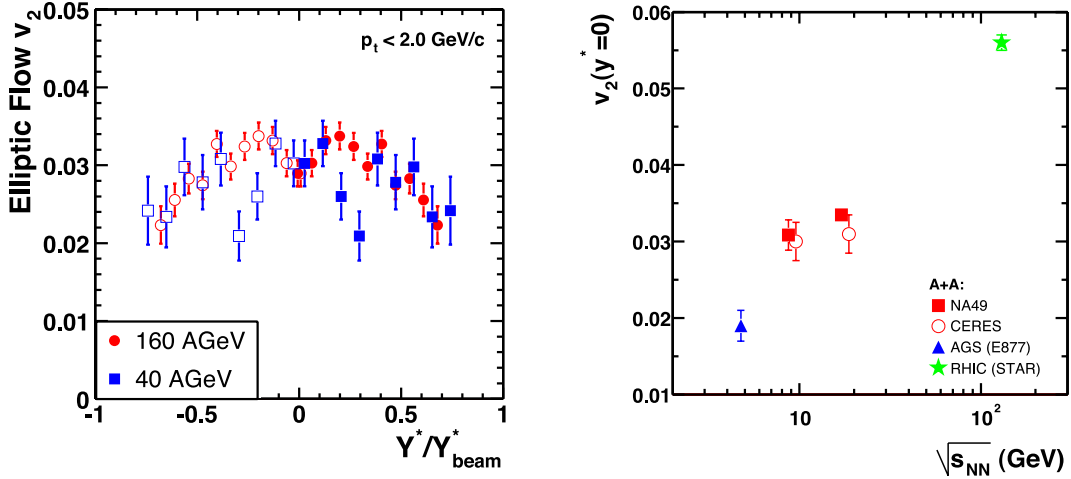}\end{center}\vspace{-1.0cm}\caption{Left: $\pi$ elliptic flow for mid-central collisions versus the center of mass rapidity in comparions for the two energies measured by NA49. Right: $\pi$ elliptic flow at midrapidity for 25\% centrality versus the center of mass energy per nucleon pair $\sqrt{s_{NN}}$ \cite{Vol00,Fil01,Adl02}.}\vspace{-0.5cm}\label{edep}\end{figure}

Due to the different coverages of phase space at the two energies, we prefer for the comparison differential to integrated values. 
As the shape of proton flow changes between the two energies and directed flow signal is small near midrapidity, we restrict the comparison to pion elliptic flow. Comparing the rapidity dependence of elliptic flow for mid-central collisions scaled to beam rapidity in the center of mass frame (Figure \ref{edep} left side), one can see a remarkable similarity of the distributions obtained from 40~AGeV and 158~AGeV data, except for two points.
To avoid problems with the different rapidity coverage of different experiments, only midrapidity values are now compared to measurements by other experiments (Figure \ref{edep} right side). 
Here one sees a rise from AGS to RHIC, but at SPS between 40 and 158~AGeV it is smaller than anticipated from the overall energy dependence.  However the size of the error bars still leaves the possibility of a constant rise. The NA49 measurements also are in good agreement with the CERES results. 

\vspace{0.4cm}
\begin{small}
Acknowledgements: This work was supported by the Director, Office of Energy Research, Division of Nuclear Physics of the Office of High Energy and Nuclear Physics of the US Department of Energy (DE-ACO3-76SFOOO98 and DE-FG02-91ER40609), the US National Science Foundation, the Bundesministerium f\"ur Bildung und Forschung, Germany, the Alexander von Humboldt Foundation, the UK Engineering and Physical Sciences Research Council, the Polish State Committee for Scientific Research (5 P03B 13820 and 2 P03B 02418), the Hungarian Scientific Research Foundation (T14920 and T23790), Hungarian National Science Foundation, OTKA, (F034707), the EC Marie Curie Foundation, and the Polish-German Foundation.
\end{small}

\end{document}